\documentclass[10pt,english,prl,twocolumn,superscriptaddress,nofootinbib,preprintnumbers]{revtex4}
\usepackage[T1]{fontenc}
\usepackage[latin1]{inputenc}
\usepackage{graphicx}

\makeatletter



\usepackage{babel}
\makeatother
\begin{document}

\title{Growing Matter}

\author{Luca Amendola}

\address{INAF/Osservatorio Astronomico di Roma, Via Frascati 33, 00040 Monte
Porzio Catone, Roma, Italy}

\author{Marco Baldi}

\address{Max-Planck-Institut f\"ur Astrophysik, Karl-Schwarzchild-Strasse
1, 85740 Garching bei M\"unchen, Germany}

\author{Christof Wetterich}

\address{Institute for Theoretical Physics, Philosophenweg 16, 69120 Heidelberg,
Germany}

\date{\today}

\begin{abstract}
We investigate quintessence cosmologies with a matter component consisting
of particles with an increasing mass. While negligible in early cosmology,
the appearance of a growing matter component has stopped the evolution
of the cosmon field at a redshift around six. In turn, this has triggered
the accelerated expansion of the Universe. We propose to associate
growing matter with neutrinos. Then the presently observed dark energy
density and its equation of state are determined by the neutrino mass. 
\end{abstract}
\maketitle
Growing observational evidence indicates a homogeneous, at most slowly
evolving dark energy density that drives an accelerated expansion
of the universe since about six billion years \cite{OBS,NB}. The
origin of dark energy is unknown, be it a cosmological constant \cite{WR},
a dynamical dark energy due to a scalar field (quintessence) \cite{WQ,RPQ},
a modification of gravity \cite{MG}, or something still unexpected.
A pressing question arises: why has the cosmological acceleration
set in only in the rather recent cosmological past? Within quintessence
models we need to explain a transition from the matter dominated Universe
to a scalar field dominated Universe at a redshift $z\simeq0.5$.

A similar crossover has happened earlier in the cosmological history,
namely the transition from radiation domination to matter domination.
This crossover is bound to happen at some time since the dilution
of the energy density with increasing scale factor $a$ obeys $\rho_{r}\propto a^{-4}$
for radiation and $\rho_{c}\propto a^{-3}$ for cold dark matter.
At some moment matter must win. We suggest in this Letter that the
presently observed crossover to a dark energy dominated Universe is
of a similar type.

We propose \char`\"{}Growing Matter\char`\"{}, an unusual form of
matter whose energy density decreases slower than the one of the usual
cold dark matter, or even increases: \begin{equation}
\rho_{g}\propto a^{3(\gamma-2)}\,,\qquad\gamma>1\,.\end{equation}
 This may be realized by particles whose mass increases with time.
In presence of both cold dark matter and growing matter a crossover
to a new epoch is then necessary at some moment. In our model this
transition is witnessed now. Similar as for the radiation-matter transition
the time for the crossover is set by the mass and abundance of the
growing matter particles. We also speculate that growing matter consists
of neutrinos. In this case the abundance is computable and the crossover
time is determined by the value of the average neutrino mass $m_{\nu}$.
Moreover, the relation between the laboratory value $m_{\nu}(t_{0})$
(for our present cosmological time $t_{0}$) and the cold dark matter
density $\rho_{c}$ at the time of the crossover only depends on dimensionless
couplings of our model.

The appearance of a substantial growing matter component strongly
influences the dynamical behaviour of the scalar field responsible
for quintessence, the cosmon. Indeed, the possibility of a time evolution
of the mass requires a time evolution of this cosmological scalar
field. In our model the mass of the particles of the growing component
obeys: \begin{equation}
m_{g}(\phi)=\bar{m_{g}}e^{-\beta\frac{\phi}{M}}\label{varying_mass}\end{equation}
 with $M\equiv1/\sqrt{8\pi G_{N}}$ the reduced Planck mass and $\bar{m_{g}}$
a constant. For $\beta<0$ an increase in $\phi$ will induce an increasing
mass.

In turn, the growing matter energy density $\rho_{g}$ influences
the evolution of the cosmon. Our approach is a model of \char`\"{}Coupled
Quintessence\char`\"{} \cite{CQ1,CQ2}. For a homogeneous cosmon field
the field equation \cite{WQ2}: \begin{equation}
\ddot{\phi}+3H\dot{\phi}=-\frac{\partial V}{\partial\phi}+\frac{\beta}{M}\rho_{g}\label{field_equation}\end{equation}
 contains a \char`\"{}force\char`\"{} $\propto\rho_{g}$ that will
counteract an increase of $\phi$ once $\beta\rho_{g}$ is comparable
to $\partial V/\partial\phi$ . In our model, this effect will eventually
dramatically slow down further evolution of the cosmon. For an almost
static $\phi(t)$ the cosmon potential $V(\phi)$ will then act similar
to a cosmological constant. The expansion of the Universe therefore
accelerates soon after $\phi$ stops to move. The coupling between
the growing matter and the scalar field ties the time of onset of
the accelerated expansion to the crossover time when $\beta\rho_{g}$
becomes important. The solution of the \char`\"{}why now\char`\"{}
problem is thus linked to the properties of growing matter. The mechanism
we propose is similar to the one presented in ref. \cite{HW}; here
however we suggest to identify the coupled matter component with the
neutrinos and we discuss the key role played by the growing matter
mass.

Let us specify our model. Besides gravity and the cosmon field, for
which we assume an exponential potential: \begin{equation}
V(\phi)=M^{4}e^{-\alpha\frac{\phi}{M}}\,,\label{field_potential}\end{equation}
 cosmology is determined by cold dark matter with a standard equation
of state $p=0$, growing matter, baryons and radiation. We denote
the fraction of homogeneous dark energy by $\Omega_{h}$, and similarly
for cold dark matter and growing matter by $\Omega_{c}$ and $\Omega_{g}$.
The cosmologically relevant parameters of our model are the dimensionless
couplings $\alpha$ and $\beta$ (eqs. \ref{varying_mass},\ref{field_potential}),
as well as the energy density of growing matter at some initial time,
e.g. $\rho_{g}(t_{eq})$ (the initial density of cold dark matter,
$\rho_{c}(t_{eq})$, can be translated to the present value of the
Hubble parameter $H_{0}$ ). We assume a flat Universe. The cosmological
equations are the standard ones, except for the modified energy-momentum
conservation for growing matter \cite{WQ2}: \begin{equation}
\dot{\rho_{g}}+3H\rho_{g}+\frac{\beta}{M}\rho_{g}\dot{\phi}=0\label{gdm_equation}\end{equation}
 which accounts for the exchange of energy between growing matter
and the cosmon \cite{WQ2,CQ1}. In case of neutrino growing matter,
eqs. (\ref{field_equation}) and (\ref{gdm_equation}) are modified
by pressure terms in early cosmology.

For the radiation and the matter dominated epochs in early cosmology
the cosmon field follows a \char`\"{}tracker solution\char`\"{} or
\char`\"{}cosmic attractor\char`\"{} with a constant fraction of early
dark energy \cite{WQ}: \begin{equation}
\Omega_{h,e}=\frac{n}{\alpha^{2}}\,,\label{early_darkenergy}\end{equation}
 where $n=3\,(4)$ for matter (radiation). This intermediate attractor
guarantees that the initial conditions for the scalar field are not
fine-tuned. Observations require that $\alpha$ is large, typically
$\alpha\ge10$ \cite{DRW}. In this \char`\"{}scaling regime\char`\"{}
one has\begin{eqnarray}
\phi & = & \phi_{0}+\frac{2M}{\alpha}\ln\left({\frac{t}{t_{0}}}\right)\,,\nonumber \\
V & \sim & \dot{\phi}^{2}\sim\rho_{c}\sim t^{-2}\,,\nonumber \\
m_{g} & \sim & \Omega_{g}\sim t^{2(\gamma-1)}\,,\quad\rho_{g}\sim t^{2(\gamma-2)}\,,\label{eq:scal1}\\
\gamma & = & 1-\frac{\beta}{\alpha}\,.\nonumber \end{eqnarray}
 The growing matter plays no role yet. Its relative weight $\Omega_{g}$
grows, however, for $\gamma>1$ or $\beta<0$ such that growing matter
corresponds to an unstable direction. The scaling regime ends once
$\gamma\Omega_{g}$ has reached a value of order one.

The future of our Universe is described by a different attractor \cite{CQ1,CQ2},
where the scalar field and the growing matter dominate, while baryons
and cold dark matter become negligible. The energy-momentum tensor
for combined quintessence and growing matter is conserved and we define
the equation of state (EOS) in the non-relativistic regime: \begin{equation}
w=\frac{p_{h}}{\rho_{h}+\rho_{g}}\,.\label{eq:eqw}\end{equation}
Notice that this is indeed the dark energy EOS measured by eg. supernovae
experiments since the two coupled fluids behave at the background
level as a single conserved component.

For this future attractor the expansion of the Universe accelerates
according to ($\gamma>3/2$):\begin{eqnarray}
H(t) & = & \frac{2\gamma}{3}t^{-1}\,,\nonumber \\
w & = & -1+\frac{1}{\gamma}\,,\label{eq:scal2}\\
\Omega_{h} & = & 1-\Omega_{g}=1-\frac{1}{\gamma}+\frac{3}{\alpha^{2}\gamma^{2}}\,.\nonumber \end{eqnarray}
For large $\gamma$ the total matter content of the Universe, $\Omega_{M}=\Omega_{c}+\Omega_{b}+\Omega_{g}$,
will be quite small in the future, $\Omega_{M}\approx\Omega_{g}\approx1/\gamma$.
The presently observed value $\Omega_{M}\approx0.25$ indicates then
that we are now in the middle of the transition from matter domination
($\Omega_{M}\approx1-3/\alpha^{2}$) to a scalar field dominated cosmology
($\Omega_{M}\approx1/\gamma$). 

The limiting case $\gamma\rightarrow\infty$ admits a particularly
simple description. In this case we encounter a sudden transition
between the two cosmic attractors at the time $t_{c}$ when the two
terms on the r.h.s. of eq. (\ref{field_equation}) have equal size,
namely for $\alpha V=-\beta\rho_{g}$ or: \begin{equation}
\Omega_{g}=\Omega_{h}/\gamma\,.\label{eq:10}\end{equation}
 While the cosmon was evolving before this time, it suddenly stops
at a value $\phi_{c}\equiv\phi(t)$ at $t_{c}$. Thus, for $t\ge t_{c}$
and large $\gamma$ the cosmology is almost the same as for a Cosmological
Constant with value $V(\phi_{c})$. On the other hand, before $t_{c}$
standard CDM cosmology is only mildly modified by the presence of
an early dark energy component (\ref{early_darkenergy}). For large
enough $\alpha$ this ensures compatibility with observations of CMB
anisotropies and structure formation. The redshift of the transition
$z_{c}$ may be estimated by equating the potential $V$ at the end
of the scaling solution (\ref{eq:scal1}) to its present value. In
terms of the present dark energy fraction $\Omega_{h,0}\approx0.75$
it is given by: \begin{equation}
\frac{H^{2}(z_{c})}{H_{0}^{2}}=\frac{2\Omega_{h,0}\alpha^{2}}{3}\,\label{z_crossover}\end{equation}
whose solution can be approximated as $1+z_{c}\approx[2\Omega_{h,0}\alpha^{2}/(3-3\Omega_{h,0})]^{1/3}$.
In the numerical examples below we will assume $\alpha=10$ and either
$\gamma=5.2$ or $\gamma=40$. Then we obtain numerically $z_{c}\approx6(5)$
for $\gamma=5.2(40)$. Thus $z_{c}$ is large enough not to affect
the present supernovae observations. The large-$\gamma$-limit is
therefore compatible with all present observations provided that $\alpha$
is large enough. We plot the time evolution of the different cosmic
components and the effective equation of state for the combined cosmon
and growing matter components in Fig. 1. For not too large $\alpha$
and $\gamma$ our model differs from $\Lambda$CDM, and we will come
back below to the interesting possibilities of observing these deviations.

So far we have made no assumptions about the constituents of the growing
matter component. It could be a heavy or superheavy massive particle,
say with a mass $1$TeV or $10^{16}$TeV. Then growing matter is non-relativistic
at all epochs where it plays a role in cosmology. In this case the
initial value $\rho_{g}(t_{eq})$ has to be chosen such that the crossover
occurs in the present cosmological epoch. Even more interesting, growing
matter could be associated with neutrinos. In this case our model
shares certain aspects with the \char`\"{}Mass Varying Neutrinos\char`\"{}
scenario \cite{MVN}, although being much closer to \char`\"{}standard\char`\"{}
Coupled Quintessence \cite{CQ2}. Neutrino growing matter offers the
interesting perspective that no new particles (besides the cosmon
and cold dark matter) need to be introduced. Furthermore, the present
value of $\rho_{g}$ can be computed from the relic neutrino abundance
and the present (average) neutrino mass $m_{\nu}(t_{0})$ (assuming
$h=0.72$): \begin{equation}
\Omega_{g}(t_{0})=\frac{m_{\nu}(t_{0})}{16eV}\,.\end{equation}
 For large $|\beta|$ the neutrino mass becomes rapidly very small
in the past such that neutrinos cannot affect the early structure
formation. The standard cosmological bounds on the neutrino mass \cite{NB}
do not apply.

For a given neutrino mass $m_{\nu}(t_{0})$ our model has only two
parameters, $\alpha$ and $\beta$ (or $\alpha$ and $\gamma$). They
will determine the present matter density $\Omega_{M}(t_{0})$. Replacing
$\gamma$ by $\Omega_{M}(t_{0})$, our model has then only one more
parameter, $\alpha$, as compared to the $\Lambda$CDM model. For
an analytical estimate of the relation between $\Omega_{M}(t_{0})=1-\Omega_{h}(t_{0})$
and $m_{\nu}(t_{0})$ we use the observation that the ratio $\Omega_{g}/\Omega_{h}$
(averaged) has already reached today its asymptotic value (\ref{eq:scal2},\ref{eq:10})
: \begin{equation}
\Omega_{h}(t_{0})=\left[\frac{\gamma}{1-\frac{3}{\alpha^{2}\gamma}}-1\right]\frac{m_{\nu}(t_{0})}{30.8h^{2}eV}\approx\frac{\gamma m_{\nu}(t_{0})}{16eV}\,.\label{eq:oh0}\end{equation}
 This important relation determines the present dark energy density
by the neutrino mass and $\gamma$: \begin{equation}
\left[\rho_{h}(t_{0})\right]^{1/4}=1.07\left(\frac{\gamma m_{\nu}(t_{0})}{eV}\right)^{1/4}10^{-3}eV.\end{equation}
 This value will change very slowly in the future since the value
$\gamma=5.22(800)$ for the maximal (minimal) neutrino mass (no sterile
neutrinos) $m_{\nu}(t_{0})=2.3eV(0.015eV)$ must indeed be large and
$w$ is therefore close to $-1$, cf. Fig. 1. The late dark energy
density is essentially determined as the neutrino energy density times
$\gamma$. Its actual value is given by the value of the scalar potential
at the crossing time $t_{c}$, i.e. $9M^{2}H^{2}(t_{c})/2\alpha^{2}$.
Since the equation of state (\ref{eq:eqw}) is today already near
the asymptotic value (\ref{eq:scal2}), cf. Fig. 1, we can relate
it to the neutrino mass ($\Omega_{h,0}\approx3/4$) by eqs. (\ref{eq:scal2},\ref{eq:oh0})\begin{equation}
w=-1+\frac{m_{\nu}(t_{0})}{12eV}\,.\label{eq:eos}\end{equation}
This remarkable expression yields $m_{\nu}(t_{0})<2.4$eV for $w<-0.8$.

How can our model based on growing matter be tested and constrained?
First of all, the presence of early dark energy manifests itself by
the detailed peak location of the CMB anisotropies \cite{D1}, the
change in the linear growth of cosmic structures \cite{FJ,D2}, and
the abundance and properties of nonlinear structures \cite{BDW}. 

Second, for not too large $\gamma$ there would be a sizable fraction
of growing matter today (for neutrino growing matter this would require
rather large neutrino masses). Then the present matter density $\rho_{M}=\rho_{c}+\rho_{b}+\rho_{g}$
differs from the (rescaled) matter density in the early Universe $\rho_{c}+\rho_{b}$.
This may affect the matching of the present values of $\Omega_{M}$
and $\Omega_{b}/\Omega_{M}$ obtained from supernovae, baryon acoustic
oscillations and clusters, with determinations from the CMB at high
redshift, through the value of $t_{eq}$ and the baryon content of
the Universe at last scattering. This effect is small for large values
of $\gamma$ (small neutrino mass).

Third, growing matter can affect the formation of structures in the
late stages. For very massive particles, growing matter would consist
of relatively few particles which have presumably fallen into the
cold dark matter structures formed in early cosmology. For scales
smaller than the range of the cosmon interaction these particles feel
a strong mutual attraction, enhanced by a factor $(2\beta^{2}+1)$
as compared to gravity. This force is mediated by the cosmon \cite{CQ1,CWVQ}.
Thus, once a sufficient $\Omega_{g}$ is reached, the growing matter
structures $\delta\rho_{g}$ grow rapidly. They will influence, in
turn, the structures in baryons and cold dark matter once the gravitational
potential of the growing matter structures becomes comparable to the
one of the cold dark matter structures. This happens rather late,
especially for large $\gamma$ since growing matter constitutes only
a small fraction of the present matter density in this case. 

The condition for the onset of an enhanced growth of $\delta\rho_{g}$
requires that the average cosmon force $\sim2\beta^{2}\Omega_{g}$
is comparable to the average gravitational force $\sim\Omega_{M}$.
This happens first at a redshift $z_{eg}$ somewhat larger that the
crossover $z_{c}$ (eq. \ref{z_crossover}). At this time the scaling
solution is still valid, with $\Omega_{M}\approx1-3/\alpha^{2}$ and
$\Omega_{g}(z)=\left[(1+z_{c})/(1+z)\right]^{3(\gamma-1)}\Omega_{g}(z_{c})$,
$\Omega_{g}(z_{c})\approx\gamma\Omega_{V}(z_{c})\approx3\gamma/(2\alpha^{2})$,
resulting in: \begin{equation}
\frac{1+z_{eg}}{1+z_{c}}=\left\{ 3\gamma(\gamma-1)^{2}\right\} ^{\frac{1}{3(\gamma-1)}}\,.\end{equation}
 For large $\gamma$ one finds $z_{eg}$ quite close to $z_{c}$ such
that the enhanced growth concerns only the very last growth epoch
before the accelerated expansion reduces further linear growth in
the dark matter component. For heavy growing matter this results in
an enhancement of $\sigma_{8}$ as compared to the $\Lambda$CDM model,
which may be compensated by a slower growth rate before $z_{c}$ due
to early dark energy \cite{D2,FJ}. 

We notice that the fluctuation growth rate $f\equiv\frac{d\log\delta}{d\log a}$
of the dominant form of matter can be estimated analytically during
both the scaling phase and the future attractor. During the first
phase the matter fluctuations grow as \cite{FJ}\begin{equation}
f_{1}=\frac{1}{4}(-1+\sqrt{24\Omega_{M}+1})\end{equation}
and are therefore slowed down with respect to the standard matter
dominated rate; for large $\alpha$ the modification is small, $f_{1}\approx1-9\alpha^{-2}/5$.
When the growing component becomes dominant, its growth rate can be
evaluated as a function of $\alpha,\gamma$ that for $\alpha\gg1$
and $\gamma\gg1$ reduces to $f_{2}\approx1.04\alpha\sqrt{\gamma}$
\cite{D2,CQ2}. Since in our model we are not yet on the final attractor
the present growth rate of the dark matter component will be quite
smaller than $f_{2}$. It is interesting to remark that in principle
an estimate of $f_{1}$ and $f_{2}$ would completely fix both parameters
$\alpha,\gamma$.

\begin{figure}
\includegraphics[scale=0.9]{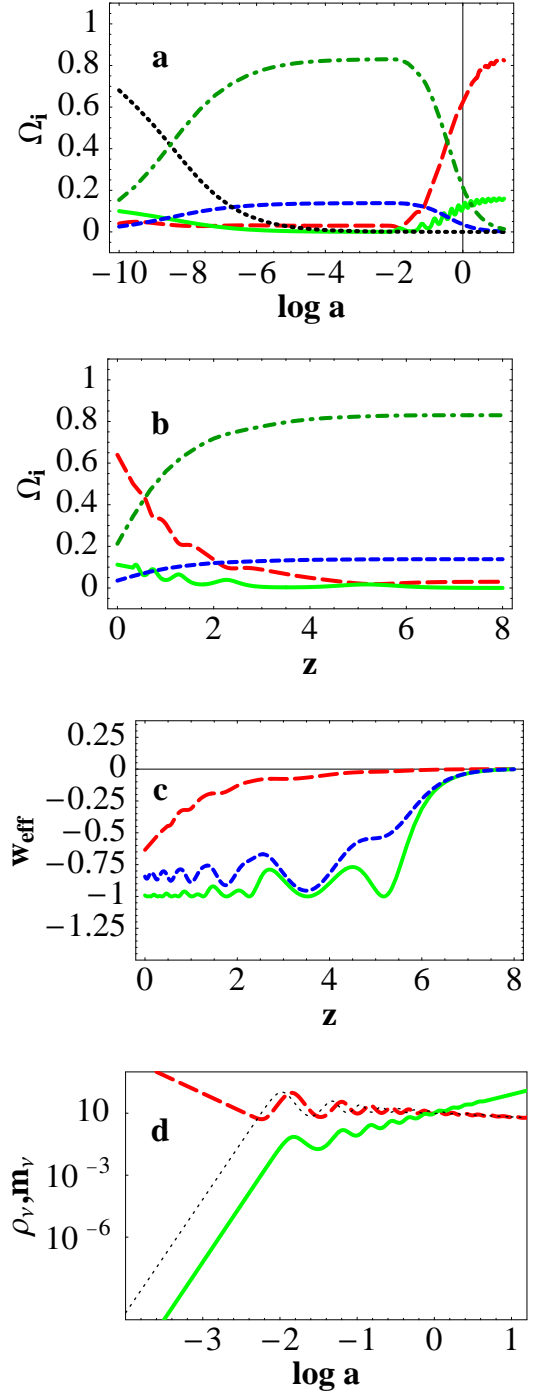}

\caption{Cosmological evolution for neutrino growing matter for $\alpha=10,\beta=-52$
and $m_{\nu,0}=2.3$eV. Panel $a$): density fractions $\Omega_{rad}$
(black, dots), $\Omega_{c}$ (dark green, dot-dashes), $\Omega_{b}$
(blue, short dashes), $\Omega_{h}$ (red, long dashes) , $\Omega_{g}$
(light green, solid). Panel $b$): blow out of panel $a$) near the
present time. Panel $c$): total equation of state $w_{eff}\equiv p_{tot}/\rho_{tot}$
(red, long dashes); combined EOS of cosmon and neutrinos (blue, short
dashes); and EOS of cosmon alone (green, solid). Panels $a)-c)$ remains
almost identical for heavy growing matter. Panel $d$): neutrino energy
density (red, long dashes), neutrino mass (green, solid) normalized
to unity today. The dotted curve represents the energy density of
always non-relativistic heavy growing matter.}
\end{figure}

An enhanced growth of $\delta\rho_{g}$ concerns only structures with
size smaller than the range $l_{\phi}$ of the cosmon-mediated interaction.
In a cosmological situation we have to solve, in principle, the coupled
system of linear fluctuation equations in $\phi$, $\rho_{g}$, $\rho_{c}$,
etc... For a rough estimate of the (density dependent) cosmon mass
we consider a fixed $\rho_{g}$, \begin{equation}
m_{\phi}^{2}=l_{\phi}^{-2}=\frac{\partial^{2}V}{\partial\phi^{2}}=\frac{\alpha^{2}V}{M^{2}}\,.\end{equation}
During the scaling solution (\ref{eq:scal1}) the cosmon range is
given by \cite{CQ1}: \begin{equation}
l_{\phi}(t)=\frac{\sqrt{2}}{3}H^{-1}(t)\,.\label{eq:lphi}\end{equation}
After $t_{c}$ the evolution of $\phi$ essentially stops, resulting
in a constant range $\hat{l}_{\phi}=\sqrt{2}/(3H(t_{c}))$. 

A different regime of growth applies for $l>l_{\phi}$. A {}``window
of adiabatic fluctuations'' opens up in the range $l{}_{\phi}<l<H^{-1}$
where the fluctuations of the coupled cosmon fluid and growing matter
can be approximated as a single fluid. In this regime the enhanced
growth is weakened by the small range of the cosmon interaction.

Neutrino growing matter was relativistic in earlier time, so that
free streaming prevents clustering. For $\beta<0$ neutrinos have
actually remained relativistic much longer than neutrinos with constant
mass. In the limit of large $\gamma$ one can estimate that the neutrinos
are relativistic at $a<a_{R}$ where\begin{equation}
a_{R}\approx(m_{\nu}(t_{0})/T_{\nu,0})^{-1/4}=0.11\left(\frac{m_{\nu}(t_{0})}{1eV}\right)^{-1/4}\label{eq:nre}\end{equation}
which corresponds to $z_{R}\in(2-10)$ for $m_{\nu,0}\in(0.015-2.3)$
(we use $m_{\nu}(a_{R})=T_{\nu}(a_{R})$, and approximate $\rho_{g}\sim const$,
so that $m_{\nu}\sim Va^{3}$, $V\sim const$, $T_{\nu}=T_{\nu,0}/a$
where $T_{\nu,0}$ is the present neutrino temperature for massless
neutrinos). The growth of neutrino structures only starts for $z<z_{R}$.
Even then, neutrinos cannot  cluster on scales smaller than their
\char`\"{}free streaming scale\char`\"{} $l_{fs}$ . This scale is
given by the time when the neutrinos become non-relativistic, eq.
(\ref{eq:nre}),  close to $H^{-1}(a_{R})\approx$ 200 $(m_{\nu}/1eV)^{3/8}$
$h^{-1}$Mpc $\in(100-1000)$ $h^{-1}$Mpc. 

On the other hand, the range $l_{\phi}$ of the effective cosmon interaction
increases, eq. (\ref{eq:lphi}) until it reaches $\hat{l}_{\phi}$.
For scales within the window $l_{fs}<l<l_{\phi}$ the neutrino clustering
is strongly enhanced (for $z<z_{eg}$) due to the additional attractive
force mediated by cosmon exchange. This enhanced clustering starts
first for scales close to $l_{fs}$ . One may thus investigate the
possible formation of lumps with a characteristic scale around $l_{fs}$.
For the range $l_{\phi}<l<H^{-1}$ one expects again an adiabatic
growth of the coupled neutrinos and cosmon fluctuations, approximated
by a single fluid. In summary, on large scales $l>l_{fs}$ the neutrino
fluctuations grow similar to the heavy growing matter fluctuations.
The growth starts, however, only very late for $z>z_{R}$ and only
from a low level given by the tiny fluctuations in a relativistic
fluid at $z_{R}$. Furthermore, neutrino fluctuations with a scale
$l<l_{fs}$ are suppressed by free streaming.

In this context one may ask to what extent the local variations of
the cosmon field $\phi$ affect the local values of the neutrino mass
and therefore the relation of cosmology to the possible outcome of
laboratory experiments. Such variations in $\phi$ are induced by
local neutrino concentrations. For neutrino lumps the perturbations
of $\phi$ are of the order of $\beta$ times the Newtonian potential
$\Phi$. Thus one expects local variations of the growing matter mass
of the order of $\Delta m_{g}/m_{g}\approx\beta^{2}\Phi$. In all
observable (non compact) astrophysical objects one has $\Phi\le10^{-4}$
and therefore it turns out that for $\beta<100$ the spatial variations
of $m_{g}$ should not be very large. In particular, we can safely
identify the cosmological value of the neutrino mass $m_{\nu}(t_{0})$
with the one measured on Earth.

A close look at Fig. 1 shows oscillations of $\Omega_{g}$, starting
around $z_{c}$ and being damped subsequently. Both the oscillation
period and the damping time can be understood in terms of the eigenvalues
of the stability matrix for small fluctuations around the future attractor
solution \cite{CQ1,CQ2}. We note, however, that the oscillations
concern only the relative distribution between $\Omega_{g}$ and $\Omega_{h}$,
while the sum $\Omega_{h}+\Omega_{g}=1-\Omega_{M}$ remains quite
smooth. A detection of the oscillations by investigations of the background
evolution, like supernovae, seems extremely hard - the luminosity
distance is a very smooth function of $z$. For neutrino growing matter
the oscillations of $m_{\nu}(t)$ around its average value induce
an uncertainty in the estimate of the relation between $\Omega_{g}(t_{0})$
and $m_{\nu}(t_{0})$ (\ref{eq:oh0}) ranging from 25\% for $\gamma=5.2$
($m_{\nu}(t_{0})=2.3$eV) to 100\% for $\gamma=40$ ($m_{\nu}(t_{0})=0.3$eV
). The relations (\ref{eq:oh0}-\ref{eq:eos}) all involve the averaged
neutrino mass.

While for large enough $\gamma$ and $\alpha$ the cosmology seems
rather realistic, one may ask if our proposal for a resolution of
the \char`\"{}why now\char`\"{} problem has not introduced other unnaturally
small parameters. Indeed, extrapolating the masses of the growing
matter particles back to the Planck time may result in extremely small
masses if $|\beta|$ is large, given typical values $\phi(t_{Pl})\approx0$,
$\alpha\phi(t_{0})/M\approx276$ (from eq. \ref{eq:scal1}). However,
we have explored here only the simplest possibility of constant $\alpha$
and $\beta$. It is well conceivable (and quite likely) that in a
fundamental theory $\alpha$ and $\beta$ are functions of $\phi$.
Slow changes will not affect our phenomenological discussion which
only concerns a rather small range of $\phi/M$. In contrast, extrapolations
back to the Planck epoch or Inflation could look completely different.
Our scenario does not need a huge overall change of the mass of the
growing matter particles. For neutrinos a growth of the mass by a
factor $10^{7}$, corresponding in the seesaw mechanism to a decrease
of the right handed neutrino mass from $M$ to $10^{11}GeV$ , would
largely be sufficient, provided a fast change happens during recent
cosmology. We also have assumed here that cold dark matter has a negligible
coupling to the cosmon. While the cosmon coupling to baryons must
be very small in order to remain compatible with the tests of General
Relativity, the coupling between cold dark matter and the cosmon is
only restricted by cosmology \cite{CQ2} and needs exploration in
our context. 

The most crucial observational issues can be understood by concentrating
on constant parameters $\alpha$, $\beta$ (and possibly a constant
cosmon-cold dark matter coupling). It will be a challenge to measure
them or to falsify the growing matter scenario. For neutrino growing
matter a determination of $\alpha$ and $\beta$ would fix the neutrino
mass, allowing for an independent test of this hypothesis by comparing
with laboratory experiments.


\begin{thebibliography}{10}
\bibitem{OBS}P. Astier et al, Astron. Astrophys. \textbf{447} 31
(2006); T. M. Davis et al., astro-ph/0701510

\bibitem{NB}D.N. Spergel et al. (WMAP collaboration) ApJS, 170, 377
(2007) arXiv:astro-ph/0603449

\bibitem{WR} S.~Weinberg, 
 Rev.\ Mod.\ Phys.\  \textbf{61} (1989) 1. 


\bibitem{WQ} C.~Wetterich, 
 Nucl.\ Phys.\  B \textbf{302} (1988) 668. 


\bibitem{RPQ}P.~J.~E.~Peebles and B.~Ratra, 
 Astrophys.\ J.\  \textbf{325} (1988) L17. 


\bibitem{MG}G.R. Dvali, G. Gabadadze and M. Porrati, Phys. Lett.
B \textbf{484}, 112 (2000)

\bibitem{CQ1}C.~Wetterich, 
 Astron.\ Astrophys.\  \textbf{301} (1995) 321 {[}arXiv:hep-th/9408025].


\bibitem{CQ2} L. Amendola, Phys. Rev. D62 (2000) 043511 {[}astro-ph/9908023],
L. Amendola, D. Tocchini-Valentini, Phys. Rev. D66 (2002) 043528 {[}astro-ph/0111535].

\bibitem{WQ2} C.~Wetterich, 
 Nucl.\ Phys.\  B \textbf{302} (1988) 645. 


\bibitem{HW}G. Huey \& B. Wandelt, Phys. Rev. D74 (2006) 023519 

\bibitem{DRW} M.~Doran, G.~Robbers and C.~Wetterich, 
 Phys.\ Rev.\  D \textbf{75} (2007) 023003 {[}arXiv:astro-ph/0609814].


\bibitem{MVN} P. Gu, X. Wang and X. Zhang, Phys. Rev. D68, 087301
(2003); R. Fardon, A.E. Nelson and N. Weiner, JCAP 0410 (2004) 005;
A. W. Brookfield et al., Phys. Rev. Lett. 96 (2006) 061301; N. Afshordi,
M. Zaldarriaga \& K. Kohri Phys. Rev. D72 (2005) 065024; O. E. Bjaelde
et al. arXiv:0705.2018; K. Ichiki \& Y. Keum, arXiv:0705.2134.

\bibitem{D1}M.~Doran, M.~J.~Lilley, J.~Schwindt and C.~Wetterich,
 Astrophys.\ J.\  \textbf{559} (2001) 501 {[}arXiv:astro-ph/0012139].


\bibitem{FJ} P.~G.~Ferreira and M.~Joyce, 
 Phys.\ Rev.\ Lett.\  \textbf{79} (1997) 4740 {[}arXiv:astro-ph/9707286].


\bibitem{D2} M.~Doran, J.~M.~Schwindt and C.~Wetterich, 
 Phys.\ Rev.\  D \textbf{64} (2001) 123520 {[}arXiv:astro-ph/0107525].


\bibitem{BDW} M.~Bartelmann, M.~Doran and C.~Wetterich, Astron.
\& Astrophys. 454 (2006) 27 
 {[}arXiv:astro-ph/0507257]. 


\bibitem{CWVQ}C. Wetterich JCAP 0310 (2003) 002 {[}arXiv: hep-ph/0203266]
\end{thebibliography}
\end{document}